\documentclass[aps,twocolumn,showkeys,groupedaddress,prb,floatfix]{revtex4-1}
\usepackage{graphicx}
\usepackage{amsmath}
\usepackage{amssymb}
\begin{document}

\title{Ion-water clusters, bulk medium effects, and ion hydration}
\author{Safir Merchant}
\author{Purushottam D. Dixit}
\author{Kelsey R. Dean}
\author{D. Asthagiri}\thanks{Corresponding author: Fax: +1-410-516-5510; Email: dilipa@jhu.edu}
\affiliation{Department of Chemical and Biomolecular Engineering and The Institute of NanoBioTechnology,  Johns Hopkins University, Baltimore, MD 21218}

\date{\today}
\begin{abstract}
Thermochemistry of gas-phase ion-water clusters together with estimates of the hydration free energy of the clusters and the water ligands are used to calculate the hydration free energy of  the ion. Often the hydration calculations use a continuum model of the solvent. The primitive quasichemical approximation to the quasichemical theory provides a transparent framework to anchor such efforts. 
Here we evaluate the approximations inherent in the primitive quasichemical approach and elucidate the different roles of the bulk medium.  We find that the bulk medium can stabilize configurations of the cluster that are usually 
not observed in the gas phase, while also simultaneously lowering the excess chemical potential of the ion. This effect is more pronounced for soft ions.  Since the coordination number that minimizes the excess chemical potential of the ion is identified as the optimal or most probable coordination number,  for such soft ions, the optimum cluster size and the hydration thermodynamics obtained without account of the bulk medium on the ion-water clustering reaction can be different from those observed in simulations of the aqueous ion.  
The ideas presented in this work are expected to be relevant to experimental studies that translate thermochemistry of ion-water clusters 
to the thermodynamics of the hydrated ion and to evolving theoretical approaches that combine high-level calculations on clusters with coarse-grained models of the medium. 
\end{abstract}  

\maketitle

\section{Introduction} \label{sec:intro}
Ion-water clusters represent the transition between an ion in the gas phase and a fully hydrated ion. Hence these systems
have been extensively studied to understand the thermochemistry of hydration and to infer how bulk properties emerge from properties of ions in clusters \cite{castleman:jpc96, jung:chemrev06, beyer:massspec07, bondy:irevpc02, robert:annrev03, kebarle:anrevpc77, niedner:chemrev00, keutsch:pnas01}. Most notably, attempts have been made to 
infer the hydration free energy of an isolated ion, a thermodynamic descriptor of the nonideal interactions between the ion and 
water, by analyzing the thermochemistry of forming ion-water clusters ${\rm X[H_2O]_n}$\cite{coe:irevpc01, klots:jpc81,coe:jpca98}, where X is the ion.  

Thermochemical data alone is inadequate for inferring the structural characteristics of hydration. In this regard, theoretical calculations
of  cluster formation  have proven useful. (For example, see Refs.\ \onlinecite{xantheas:jacs95, novoa:jpca97, topol:jcp99, zhan:jcpa01, pliego:jcpa01,brya:jpcb08, kelly:jpcb06, busta:jpca11}.) Moreover, such calculations together with hydration free energy calculations of the clusters and water ligands have been used for estimating the hydration free energy of single ions. In calculating the hydration free energies, almost always a continuum model for the bulk solvent is assumed and either one \cite{pliego:jcpa01} or an ensemble 
\cite{topol:jcp99} of gas phase configurations of the cluster is used for obtaining the hydration free energy of the cluster. In some studies the gas-phase cluster is also allowed to relax in response to the model bulk medium \cite{zhan:jcpa01}.

An important development in the statistical mechanics of hydration has been the quasichemical organization \cite{lrp:mp98,lrp:ES99,pauli:advpc02,lrp:book,lrp:cpms} of the potential distribution theorem. This approach provides a rigorous, statistical mechanical framework to relate the thermodynamics of solute-water clusters to the bulk hydration of the solute. In practical implementations, 
approximations to the quasichemical theory are invariably made. In the {\em primitive} quasichemical approximation,  --- the adjective {\em primitive} indicates neglecting the role of the bulk medium on the local clustering reaction --- the thermochemistry of ion-water clustering in the ideal gas phase, obtained using standard quantum chemistry approaches, is coupled with estimates of the response
of the bulk medium, treated either as a dielectric continuum \cite{rempe:jacs00,rempe:fpe01,grab:jpca02, asthagiri:cpl03,asthagiri:cpl032, rempe:pccp04} or as a discrete molecular solvent \cite{asthagiri:jcp03}.

Within primitive quasichemical theory, the hydration free energy is estimated by varying the number of water molecules 
in the cluster and finding the optimal coordination number that minimizes the hydration free energy of the ion.  This cluster variation approach has proven successful in establishing the hydration structure of small, hard ions such as  Li$^+$(aq) \cite{rempe:jacs00}, Na$^+$(aq) \cite{rempe:fpe01}, and Be$^{2+}$(aq) \cite{asthagiri:cpl03}; for these cases, the optimal coordination structure predicted by primitive quasichemical theory is in good agreement with the most probable coordination observed in {\em ab initio\/} molecular dynamics simulations (AIMD).  Similar agreement is also seen for Mg$^{2+}$, Ca$^{2+}$, and some of the transition metal dications \cite{lrp:zn04}. Given the then limitations of small systems, short simulation times, and uncertain quality of the underlying electro density functionals in 
AIMD simulations (for example, Refs.\ \onlinecite{asthagiri:waterpre,schwegler:jcp04a}), primitive quasichemical theory proved  useful in cross-checking the simulation results itself.  (The predictions for H$^+$\cite{asthagiri:hpnas} and HO$^-$\cite{asthagiri:hopnas} are also in fair agreement with AIMD simulations, but consensus remains elusive\cite{voth:acc06,tuckerman:acc06}. These systems also challenge both theory and simulations because of the need to describe nuclear quantum effects.)

For the soft K$^+$ ion, primitive quasichemical predicts an optimum coordination with four (4) water molecules \cite{rempe:pccp04}. But the results from AIMD simulations are less conclusive.  Some studies identify an inner-coordination number of four and a second outer-population with two additional water molecules within the nominal first hydration shell of the ion  \cite{rempe:pccp04,rempe:bc06}, but
others make no such distinction \cite{whitfield:jctc07}. Beside K$^+$,  it has also been well appreciated that for some of the halides, optimum gas-phase clusters can show intermolecular bonding between the coordinating shell water molecules \cite{topol:jcp99,xantheas:jpc96}, a feature that is not usually observed in the coordination structure for the hydrated ion in the liquid. Thus the primitive quasichemical approach for these ions has not been successful in reproducing the hydration structure and thermodynamics.  The limitations in describing the hydration of soft cations and anions suggests that the bulk medium plays an important role in the hydration structure of these ions. Developing the framework to understand this effect is the  objective of this article.

In Section~\ref{sc:theory}, we present the quasichemical theory, elucidate the role of the medium, and highlight the
physical consequences of the primitive quasichemical approximation.  In our previous work \cite{merchant:jcp09} 
we showed the importance of occupancy number variations of water in an empty coordination sphere for understanding an 
ion's coordination structure. Developing those ideas further, here we find that the bulk medium promotes a better packing of solvent molecules around the ion, leading to a decrease in the contribution to the hydration free energy due to 
interactions between the ion and the solvent molecules within the coordination sphere. The medium stabilizes configurations which are otherwise not observed in isolated ion-solvent clusters. Without proper account of this effect, the predicted optimal coordination state is typically lower than the optimal coordination observed in simulations, and appreciating this effect is necessary in calculating hydration thermodynamics as well. 

\section{Theory}\label{sc:theory}
\subsection{Quasichemical theory}
We present only the main elements of the quasichemical theory; more extensive  discussions are available elsewhere \cite{lrp:ES99,pauli:advpc02, lrp:book,lrp:cpms}. We define the coordination sphere of radius $\lambda$ around the ion, $\alpha$, 
and by so doing, separate the local, chemically intricate ion interactions with water molecules (the solvent ligands) within the coordination sphere from the interaction of the ion with the medium (the bulk) outside the coordination shell.  The chemical potential can then
be written as  \cite{lrp:ES99,pauli:advpc02, lrp:book,lrp:cpms}
\begin{eqnarray}
\beta \mu^{ex}_{\alpha}&=&  \ln x_0  -\ln p_0 + \beta \mu^{ex}_{\alpha}(0) \, ,
\label{eq:quasichemical}
\end{eqnarray} 
where $\beta = 1/k_BT$, $k_B$ is the Boltzmann constant and T is the temperature. $x_0$ is the probability of finding zero (0) solvent ligands within $\lambda$. (The position of the water molecule is defined by the position of the oxygen atom. The ion in the $n^{\rm th}$ hydration state has $n$-ligands within $\lambda$.) The contribution  to the free energy due to ion-bulk  interactions 
when the ion is in the $n=0$ state is $\mu^{\rm ex}_{\rm \alpha} (n=0) \equiv \mu^{\rm ex}_{\alpha}(0)$. The probability 
of finding zero (0) solvent ligands within $\lambda$ when the ion-water interactions are turned off is $p_0$. This factor accounts
for the free energy of creating an empty coordination sphere in the solvent and thus accounts for the packing (steric) contributions to $\mu^{\rm ex}_\alpha$. 

The quasichemical form is obtained by noting that $x_0$ is specified by chemical equilibria of the form \cite{lrp:ES99,pauli:advpc02, lrp:book,lrp:cpms,paliwal:jcp06,merchant:jcp09}
\begin{eqnarray}
{\rm \alpha[H_2O]}_0({\rm aq}) + n{\rm H_2O}({\rm aq})  &\rightleftharpoons& {\rm \alpha[H_2O]}_n({\rm aq})  \, ,
\label{eq:ionaqcoord}
\end{eqnarray} 
where $\alpha[{\rm H_2O}]_n$ denotes the ion plus $n$-water cluster within the coordination sphere and `aq'  indicates
 that the clustering reaction is in the presence of the bulk medium.  Given the equilibrium constant of the above reaction is
$K_n$ and the density of water molecules is $\rho_w$, we immediately obtain  \cite{lrp:ES99,pauli:advpc02, lrp:book,lrp:cpms,paliwal:jcp06,merchant:jcp09}
\begin{eqnarray}
 \ln x_0 &=& -\ln (1+\sum_{n \geq 1}K_n \rho_{w}^n) \, .
\label{eq:lnx0}
\end{eqnarray}  
Note that the excess chemical potential of the cluster $\alpha[{\rm H_2O}]_0 $  is just 
\begin{eqnarray}
 \beta \mu^{ex}_{\alpha\cdot w_0} &=&  \beta \mu^{ex}_{\alpha}(0) -\ln p_0 \, .
\label{eq:hardion}
\end{eqnarray}  
(For ease in writing, we will denote ${\rm H_2O}$ by $w$ in all subscripts.)

Approximating Eq.~\ref{eq:lnx0} by its maximum term, we get 
\begin{eqnarray}
\beta \mu^{ex}_{\alpha} &\approx& -\ln K_n \rho_w^n +  \beta \mu^{ex}_{\alpha\cdot w_0} \, .
\label{eq:qcapprox}
\end{eqnarray} 

$K_n$  is given by the ratio of partition functions of the cluster to individual molecules and can be challenging to calculate 
with full account of the surrounding medium \cite{pauli:advpc02,lrp:book,pratt:mp98,merchant:jcp09,lrp:jpcb01,lrp:hspre}.  But 
the equilibrium constant $K_n^{(0)}$ of the clustering reaction 
\begin{eqnarray}
{\rm \alpha[H_2O]}_0({\rm g}) + n{\rm H_2O}({\rm g})  &\rightleftharpoons& {\rm \alpha[H_2O]}_n({\rm g})  \, ,
\label{eq:iongcoord}
\end{eqnarray} 
in the ideal gas phase is more amenable to theoretical calculations. In Eq.~\ref{eq:iongcoord},  `g'  indicates that the solvent outside the coordination shell is non-interacting (or ideal); that is, the reaction is performed in the ideal gas phase. Since the free energy change 
 for Eqs.~\ref{eq:ionaqcoord} and~\ref{eq:iongcoord}, with appropriate choice of standard concentrations, are just $-k_{\rm B}T \ln K_n$
 and $-k_{\rm B}T \ln K_n^{(0)}$, respectively, we find that 
\begin{eqnarray}
k_BT\ln\frac{K_n}{K_n^{(0)}} &=& \mu^{ex}_{\alpha\cdot w_0}   -  \mu^{ex}_{\alpha\cdot w_n} + n \mu^{ex}_{w} \, .
\label{eq:knbykn0}
\end{eqnarray} 
The first two terms on the right hand side represent the hydration free energies of the clusters containing 0 waters and $n$ waters,
respectively, and $ \mu^{ex}_{w}$ is the hydration free energy of a water molecule.

A similar development can be pursued when the coordination shell is empty, which is equivalent to 
the case when the ion and solvent do not interact. Denoting the noninteracting solute by $\bullet$, the analog of Eq.~\ref{eq:knbykn0}  is
\begin{eqnarray}
k_BT\ln\frac{\tilde{K}_n}{\tilde{K}_n^{(0)}} &=& \mu^{ex}_{\bullet \cdot w_0}   -  \mu^{ex}_{\bullet \cdot w_n} + n \mu^{ex}_{w} \,.
\label{eq:knbykn0t}
\end{eqnarray} 
Note that $p_0$ is related to $\tilde{K}_n$ \cite{pauli:advpc02,paliwal:jcp06} exactly like $x_0$ is related to $K_n$ (Eq.~\ref{eq:lnx0}), and equilibria analogous to Eqs.~\ref{eq:ionaqcoord} and~\ref{eq:iongcoord} specify $\tilde{K}_n$ and $\tilde{K}^{(0)}_n$, respectively.  

From Eqs.~\ref{eq:qcapprox} and ~\ref{eq:knbykn0} we obtain:
\begin{eqnarray}
 \mu^{ex}_{\alpha} &\approx&  -k_BT\ln K_n^{(0)} \rho_w^n  +  \mu^{ex}_{\alpha\cdot w_n} -  n \mu^{ex}_{w}
\label{eq:qcapprox2}
\end{eqnarray}
 Eq ~\ref{eq:qcapprox2} not only permits the calculation of $\mu^{\rm ex}_{\alpha}$, but it also suggests an approach  to  identify the optimal coordination state. For different $n$, we can compute $K_n^{(0)}$; often highly descriptive forcefields, including
{\em ab initio} potentials, are used to model the thermochemistry of the gas-phase reaction.  The presence of the bulk medium outside the coordination shell is then corrected {\em a posteriori} by adding the free energy  of hydrating the cluster  and subtracting the free energy of transferring the requisite number of water molecules to the gas phase from the liquid. A continuum or dielectric model of the solvent is often assumed for this purpose. The coordination state that minimizes $\mu^{ex}_{\alpha}$ is then identified as the optimal coordination state \cite{asthagiri:cpl03,asthagiri:cpl032, rempe:pccp04, rempe:jacs00,rempe:fpe01,grab:jpca02}.  In the {\em primitive} quasichemical approximation, the configuration of the gas phase cluster is not allowed to relax in computing the hydration free energy. To better understand this approximation, we first identify the role of the bulk medium on local ion-water clustering. 
  
\subsection{Role of the bulk medium on local ion-water clustering}
 
The ratio of equilibrium constants $K_n$ and $\tilde{K}_n$ was earlier\cite{merchant:jcp09} shown  to be
\begin{eqnarray}
\frac{K_n}{\tilde{K}_n} &=& e^{-\beta W_n} = e^{-\beta (\mu^{ex}_{\alpha}(n) - \mu^{ex}_{\alpha}(0))}  \,,
\label{eq:mun}
\end{eqnarray}   
where $W_n$ is the free energy of forming the ion-water cluster in the presence of the bulk and $\mu^{ex}_{\alpha}(n)$ is hydration free energy of the ion in its $n^{th}$ coordination state. Specifically, 
\begin{eqnarray}
e^{-\beta \mu^{ex}_{\alpha}(n)} &=& \langle e^{-\beta \varepsilon} | n \rangle_0 \, ,
\label{eq:mun1}
\end{eqnarray}   
where $\varepsilon$ is the interaction energy of the ion with the rest of the medium, $\langle \ldots\rangle_0$ indicates averaging when the ion and the medium are thermally uncoupled (denoted by the subscript 0), and $\langle \ldots|n\rangle_0$ indicates that only cases with exactly $n$ solvent molecules within the coordination shell are considered in averaging.

By parsing the interaction energy $\varepsilon$ into a local piece, $\varepsilon_{\rm local}$, obtained by considering ion-interactions with the $n$ coordinating water molecules, and the remaining long-range piece, $\varepsilon_{\rm lr}$, we have 
\begin{eqnarray}
e^{-\beta \mu^{ex}_{\alpha}(n)} &=& \langle e^{-\beta \varepsilon_{\rm local}} | n \rangle_0 \cdot \frac { \langle e^{-\beta \varepsilon_{\rm lr} }e^{-\beta\varepsilon_{\rm local}} | n \rangle_0 } {  \langle e^{-\beta \varepsilon_{\rm local}} | n \rangle_0 } \nonumber \\
& = & \langle e^{-\beta \varepsilon_{\rm local}} | n \rangle_0 \cdot  \langle e^{-\beta \varepsilon_{\rm lr}} | n \rangle_{{\rm local}}
\label{eq:munint}
\end{eqnarray}   
where we have made use of the rule of averages \cite{pauli:advpc02, lrp:book,lrp:cpms}, in rewriting the second factor on the right; $\langle \ldots | n\rangle_{{\rm local}}$ indicates averaging such that the $n$-coordination shell molecules are thermally coupled with the ion, and the bulk medium is uncoupled 
from the ion but is coupled with the $n$-water molecules. 

Identifying the first factor on the right in Eq.~\ref{eq:munint} by $e^{-\beta \xi_{\rm aq}(n)}$ and the second factor by $e^{-\beta \mu^{ex}_{\rm outer}(n)}$, we have 
\begin{eqnarray}
e^{-\beta \mu^{ex}_{\alpha}(n)} &=& e^{-\beta \xi_{\rm aq}(n)} e^{-\beta \mu^{ex}_{\rm outer}(n)} \, .
\label{eq:mundecomp}
\end{eqnarray}     
We emphasize that $e^{-\beta \xi_{\rm aq}(n)}$ accounts for the contribution to the chemical potential due to the interaction of the ion with solvent ligands within the coordination shell {\em in the presence of the medium outside\/}, and $e^{-\beta \mu_{\rm outer}(n)}$ accounts for the contribution 
due to the interaction of the bulk medium with the ion when there are $n$ solvent ligands present inside the coordination shell. Note that 
when $n = 0$, $\mu_{\rm outer}(0) = \mu^{ex}_{\alpha}(0)$ (Eq.~\ref{eq:quasichemical}).  

Thus we find that  
 \begin{eqnarray}
\frac{K_n}{\tilde{K}_n} &=& e^{-\beta W_n} = e^{-\beta \xi_{\rm aq}(n)} e^{-\beta ( \mu^{ex}_{\rm outer}(n) -\mu^{ex}_{\alpha}(0))}  \, .
\label{eq:KnbyKnt1}
\end{eqnarray}   
(Our earlier work\cite{merchant:jcp09} had an error in factoring $K_n/\tilde{K}_n$. The error is discussed in Appendix~\ref{sc:appA} for completeness.)  

 Now if the medium outside the coordination shell was an ideal gas,
\begin{eqnarray}
\frac{K_n^{(0)}}{\tilde{K_n}^{(0)}} &=& e^{-\beta \xi_{\rm g}(n)} \, .
\label{eq:KnbyKnt2}
\end{eqnarray}  
From Eq.~\ref{eq:KnbyKnt1} and~\ref{eq:KnbyKnt2}, we get
 \begin{eqnarray}
\frac{K_n/K_n^{(0)}}{\tilde{K}_n/\tilde{K}_n^{(0)}} &=& e^{-\beta ( \mu^{ex}_{\rm outer}(n) -\mu^{ex}_{\alpha}(0))} e^{-\beta(\xi_{\rm aq}(n) - \xi_{\rm g}(n))}\, , \nonumber \\
& & 
\label{eq:KnbyKnt3}
\end{eqnarray}
and this together with Eqs.~\ref{eq:knbykn0},~\ref{eq:knbykn0t}, and \ref{eq:KnbyKnt3} gives
\begin{eqnarray}
 \mu^{ex}_{\alpha\cdot w_n} = \xi_{\rm aq}(n) - \xi_{\rm g}(n) + \mu^{ex}_{\rm outer}(n) + \mu^{ex}_{\bullet \cdot w_n} 
 \label{eq:frozen}
\end{eqnarray}
Finally, substituting in Eq.~\ref{eq:qcapprox2} and noting Eq.~\ref{eq:hardion}, we get
\begin{eqnarray}
 \mu^{ex}_{\alpha} &=& -k_BT\ln K_n^{(0)} \rho_w^n  + \xi_{\rm aq}(n) - \xi_{\rm g}(n) \nonumber \\
 & + & \mu^{ex}_{\rm outer}(n) + \mu^{ex}_{\bullet \cdot w_n}   -  n \mu^{ex}_{w}   
 \label{eq:muex}
\end{eqnarray}

\subsection{Primitive quasichemical aproximation}

Equation~\ref{eq:frozen} clearly identifies three different roles of the solvent medium and also helps us induce the physical consequences of the {\em primitive} quasichemical approach. The quantity $\xi_{\rm aq}(n) - \xi_{\rm g}(n)$ is the change in the free energy
due to the local ion solvent interaction upon coupling the gas-phase cluster with the bulk medium. 
 It accounts for the energetics associated with the relaxation of the cluster  in the presence of the medium. Utilizing gas phase geometries completely ignores this effect. Since the configuration of clusters with a high coordination state can be expected to be more sensitive to the presence of the bulk medium than configurations with  lower coordination states,  $\xi_{\rm aq}(n) - \xi_{\rm g}(n)$ is expected to be larger for higher coordination states than for lower ones.
 
The quantity $\mu^{ex}_{\rm outer}(n)$ accounts for the interaction of the bulk medium with the ion when there are $n$ solvent ligands inside the coordination shell. Since the configuration of the cluster will change in the presence of the bulk medium, 
this quantity is also inadequately described when we do not allow the gas-phase cluster to relax. We do expect, however, that 
since only long range ion-solvent interactions contribute to $\mu^{ex}_{\rm outer}(n)$, this quantity can be well captured by dielectric models of hydration. 

The final term is the hydration free energy of an $n$-water cluster without the ion. It consists of packing interactions which are also typically ignored in the {\em primitive} quasichemical approach.

\section{Methods} 
 Liquid water and ion water systems are studied under NVT conditions using Metropolis Monte Carlo simulations\cite{metropolis:jcp53, allen}. The cubic simulation cell comprises 306 water molecules for  the pure water system; ion-water system consists of an additional ion which is held fixed at the center. The box volume ($L^3$) is adjusted such that the number density in each case is 33.33 nm$^{-3}$. Water is modeled  with the SPC/E potential\cite{berendsen:jpc87}. Ion parameters and the magnitude of coordination radii are taken  from our earlier study\cite{merchant:jcp09}.  Electrostatic interactions are modeled by the generalized reaction field (GRF) approach\cite{hummer:jpc96, hummer:molphys92, hummer:physreve94,hummer:jpcond94, lrp:ionsjpca98}. Both Lennard-Jones and electrostatic interactions were truncated at $L/2$. 
 
As in our earlier work\cite{merchant:jcp09}, Na$^+$, K$^+$, and F$^-$ ions were studied. Since the trends for F$^-$ follow those of the smaller Na$^+$ ion, for clarity we only present results comparing Na$^+$ and K$^+$. The parameters for the ions were obtained from Ref.\ \onlinecite{hummer:jpc96}. 
 
 Simulations are carried out for 6$\times$10$^5$ sweeps. Each sweep consists of one attempted translation or rotation of each water molecule. The first 3$\times$10$^5$ are set aside for equilibration. During the equilibration phase the maximum angular deflection and linear displacement of the water molecule is optimized to yield an acceptance ratio of 0.3. The optimized values are then held fixed for the next 3$\times$10$^5$ sweeps. Configurations of the system are saved every 10 sweeps for analysis. 
 
 A similar simulation strategy was used for simulating ion-water and neat water clusters in the absence of the bulk medium. The 
 solvent ligands comprising the cluster are restricted to a sphere of radius $\lambda$. Clusters were simulated for 6$\times$10$^7$ sweeps with the first 3$\times$10$^7$  sweeps used for equilibration and the next 3$\times$10$^7$ sweeps used for production.  During the production phase configurations of the clusters were saved every 10 sweeps for analysis. 
    
\subsection{Free energy of forming an ion-water cluster in the presence of the bulk medium} 

We calculate $\xi_{\rm aq}(n)$, the free energy of inserting an ion within a cluster in presence of the external solvent,
 for $n = 1-6$ coordination states. For the ions studied here, the local chemical contribution $k_{\rm B}T \ln x_0$ is fully accounted 
 for by this coordination number\cite{merchant:jcp09} and thus these are the states of most interest. 
  
 $\xi_{\rm aq}(n)$ is obtained indirectly by calculating the response of the medium, $\mu^{ex}_{\rm outer}(n)$,  and subtracting it  from $\mu^{ex}_{\alpha}(n)$, the hydration free energy of the ion in the $n^{th}$ coordination state (Eq. \ref{eq:mundecomp}). Values of $\mu^{ex}_{\alpha}(n)$ were taken from our earlier work \cite{merchant:jcp09}. To obtain $\mu^{ex}_{\rm outer}(n)$, we simulate an ion water system in which the number of solvent molecules within the coordination shell was held at $n$. Further,  solvent molecules within the coordination sphere are fully-coupled with the ion, but the ion interacts with the bulk medium at various fractional charge states $q\zeta$, where $q$ is the charge on the ion and $\zeta = 0, 1/2 \pm \sqrt{12}, 1$. Ion-solvent pair correlation function was obtained
 from the $\zeta = 1$ simulation.

 The electrostatic contribution to  $\mu^{ex}_{\rm outer}(n)$ is determined by integrating the average potential, $\langle \phi \rangle_{q\zeta}$, at the centre of the ion due to the bulk medium. $q\zeta$ is the charge seen by the bulk water molecules. (We emphasize that 
 the water molecules inside the coordination sphere always see a charge of $q$.) The electrostatic contribution to $\mu^{ex}_{\rm outer}(n)$  is approximated by the two point Gauss-Legendre quadrature
 \begin{eqnarray}
 \mu^{ex}_{\rm outer, elec}(n) &=& \frac{q}{2}(\langle \phi \rangle_{1/2 + 1/\sqrt{12}} + \langle \phi \rangle_{1/2 - 1/\sqrt{12}}) \nonumber
\end{eqnarray} 
which is exact to fourth order in perturbation theory\cite{hummer:jcp96}. 

For obtaining the van der  Waals contribution to $\mu^{ex}_{\rm outer}(n)$, we determine the distribution of ion-bulk interaction energies
from the $\zeta = 0$ simulation. The van der Waals contribution to  $\mu^{ex}_{\rm outer}(n)$ is then obtained by 
approximating this distribution by a Gaussian \cite{Asthagiri:2007p323}  within the inverse form of the potential distribution theorem.

\subsection{Free energy of forming an ion-water cluster in the absence of the bulk medium} 

Ion-water clusters within the coordination sphere were studied at different fractional charge states to obtain the electrostatic contribution to the free energy of forming an ion-water cluster when there is no bulk medium outside the coordination sphere. (Ion-solvent pair correlation function was obtained from the $\zeta = 1$ simulation.) To calculate the van der Waals contribution to the free energy, we first obtain the distribution of ion-solvent interaction energies for the neutral ion. We obtain the uncoupled binding energy distribution by performing test particle insertions of the neutral ion in neat water clusters. The vdW contribution is then obtained by histogram overlap \cite{bennett:jcp76}. 

\section{Results and Discussion}

\subsection{Pair correlation in ${\rm K[H_2O]_6}$(aq) and ${\rm K[H_2O]_6}$(g) clusters}

\begin{figure}[h!]
 \includegraphics{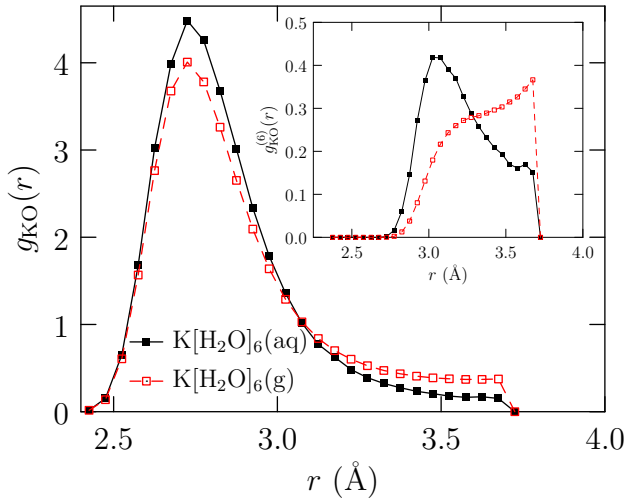}
\caption{The distribution, $g_{KO}(r)$, of water around K$^+$ in the presence (aq) and absence (g) of the external medium for
the $\rm {K[H_2O]}_6$ cluster. The radius of the coordination sphere $\lambda$ = 3.7 \AA.  The pair correlation in presence of the medium shows enhanced structure relative to that for the cluster in the absence of the bulk. This suggests that the external medium promotes a better packing of solvent around K$^+$.   {\bf Inset}: The distribution, $g_{KO}^{(6)}$, of the farthest water molecule from K$^+$. 
Observe that in the absence of the bulk medium, this water molecule is closer to the boundary of the cluster.  } \label{fg:grKW6} 
\end{figure}

For hydrated ions, an implicit assumption in using gas-phase clusters to calculate solution phase thermodynamic properties is 
that the average distribution of water molecules in the solution phase and gas phase clusters are similar.  Examining the 
ion-water pair correlation (Fig.~\ref{fg:grKW6}) does validate this assumption, but {\em only} for the low coordination states. 

Figure~\ref{fg:grKW6} shows that  the presence of the medium causes the cluster to be better packed, resulting in a somewhat sharper pair correlation for the cluster in the bulk than in its absence.  As anticipated, the effect of the medium is most pronounced
for the water molecule that is farthest from the ion (Fig.~\ref{fg:grKW6}, Inset). 
 
The distribution of the water molecules comprising the $n=1,\ldots,3$ coordination states is insensitive to the presence of the external medium (data not shown), a feature that is in accordance with our earlier \cite{merchant:jcp09} finding that there is a core group
of water molecules with which the ion interacts strongly enough that the effect of the bulk medium on these water molecules is
small.  The higher coordination states are however influenced by the bulk medium as well.  For the cluster with no bulk medium
outside, the 6$^{\rm th}$ water molecule is closer to the boundary of the cluster, a position that also allows this water molecule
to associate with the remaining water molecules in the cluster rather than with the ion. Similar incomplete shell effects, where the second hydration shell starts to form before the first shell is complete, have been inferred on the basis of experimental studies on gas phase clusters with water\cite{maut:jpc86} or ammonia\cite{searles:jpc68, castleman:jacs78} as solvent ligands. 

\begin{figure}
\includegraphics{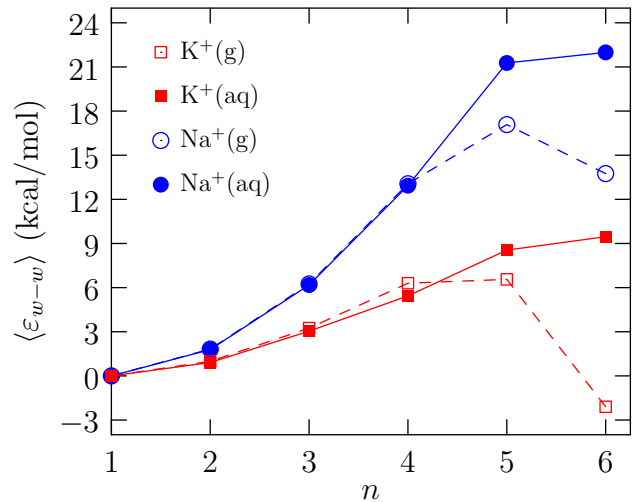}
\caption{The average solvent-solvent interaction energy, $\langle \varepsilon_{w-w}\rangle$,  of ion-water clusters for different coordination states. For curves denoted by `aq' (filled symbols), $\langle \varepsilon_{w-w}\rangle$ is obtained by extracting the cluster from simulations in the presence of the bulk medium. Curves denoted by `g'  (open symbols) correspond to clusters simulated in the 
absence of a bulk. The coordination radius $\lambda$ = 3.7 \AA. 
For higher coordination states, the average excess internal energy is lower in the case of a cluster without a bulk medium outside.}\label{fg:ions_ww} 
\end{figure}

Figure~\ref{fg:ions_ww} shows that the bulk can stabilize configurations of the water molecules that would be unfavorable otherwise. 
The average excess internal energy $\langle \varepsilon_{w-w}\rangle$ of the $n$-water molecules in the coordination sphere 
for the higher coordination states, relative to clusters in the absence of the bulk medium, 
is higher for the cluster that is extracted from simulations including the bulk.  Further, $\langle \varepsilon_{w-w}\rangle$ is relatively insensitive to the presence of the bulk for the low coordination states. Thus we find that the medium promotes better packing of water molecules around the ion by stabilizing configurations of the cluster which are otherwise unfavorable. 

\subsection{Medium effects on the free energy of forming ion-water clusters}

$W_n$, the free energy of forming an ion plus $n$-water molecule cluster comprises a local contribution, $\xi_{\rm aq}(n)$,
and the response of the bulk medium, $ \mu^{ex}_{\rm outer}(n) -\mu^{ex}_{\alpha}(0)$  (Eq.~\ref{eq:KnbyKnt1}). 
We next consider each of these contributions separately. 
 
\begin{figure}
\includegraphics{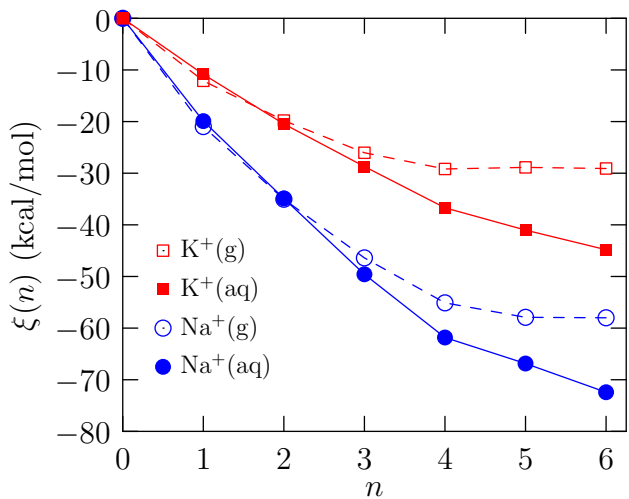}
\caption{The local contribution, $\xi (n)$, to the free energy of forming an ion plus $n$-water molecule cluster.  The filled symbols, 
$\xi_{\rm aq}(n)$, are results in the presence of the bulk (aq) and the unfilled symbols, $\xi_{\rm g}(n)$,  in its absence (g). The radius of the coordination sphere is  $3.7$ \AA. For  $n = 1,2,3$, $\xi_{\rm aq}(n)$ is insensitive to the presence of the bulk medium outside the coordination shell,  whereas for $n = 4,5,6$, $\xi (n)$ is more favorable with the medium than without.}\label{fg:xi_ions} 
\end{figure}

Figure~\ref{fg:xi_ions} shows that $\xi_{\rm aq}(n)$ is insensitive to the presence of the medium for small clusters ($n = 1,2,3$). However for larger clusters ($n = 4,5,6$), not having a bulk medium to stabilize the cluster leads to a higher free energy.  The difference, 
$\xi_{\rm aq}(n) - \xi_{\rm g}(n)$ is always negative. Thus ignoring this difference while implementing cluster variation will 
predict a less favorable contribution due to that coordination state: in effect, the probability of observing these higher coordination 
states will be predicted to be lower than when the difference is included. 

\begin{figure}
\includegraphics{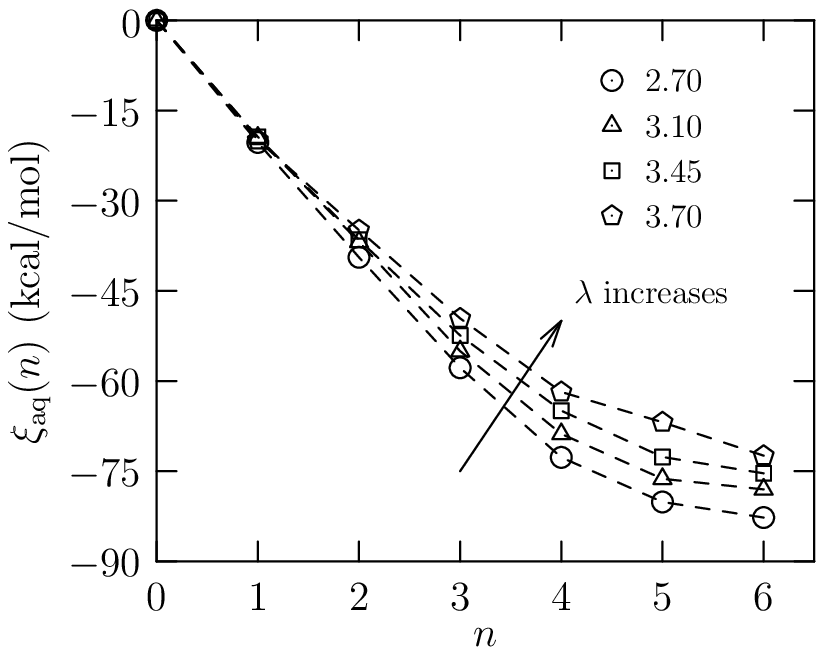}
\includegraphics{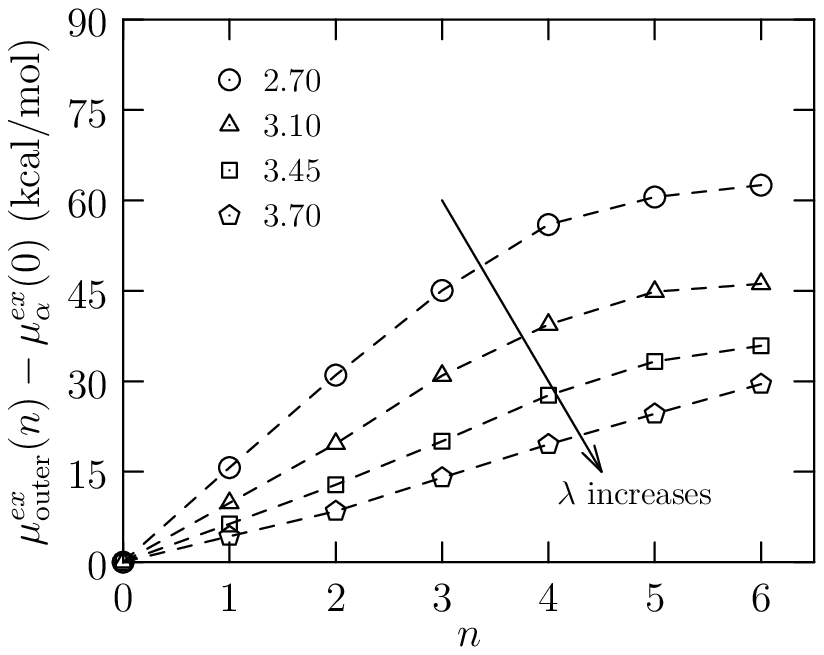}
\caption{ {\bf Top}: The local contribution, $\xi_{\rm aq}(n)$, of forming Na$^+$ plus $n$-water clusters. 
{\bf Bottom}: The long-range contribution, $\mu^{ex}_{\rm outer}(n) -\mu^{ex}_{\alpha}(0)$, to the free energy of forming Na$^+$ plus $n$-water clusters.  Note that increasing the radius $\lambda$ of the coordination sphere does not greatly affect $\xi_{\rm aq}(n)$ for $n \leq 3$ but increases  it for $n = 4,5,6$. On the same scale, the response of the medium is more pronounced for the same change in $\lambda$
and it tends to become insensitive to the presence of the coordinating solvents for large $\lambda$.}\label{fg:na_cn} 
\end{figure}  
Our previous work showed that increasing the coordination radius $\lambda$ decreases the chemical term $W_n$. 
Intuitively, we expect a more favorable clustering free energy with increasing coordination radius because the 
ion can be better accommodated by the coordination shell solvent molecules, that is the local contribution becomes favorable and
we expect $W_n$ to decrease because $\xi_{\rm aq}(n)$ is expected to decrease. However,  we find that increasing the coordination radius has no effect on the local contribution, $\xi_{\rm aq}(n)$, for small ($n \leq 3$)  clusters and \emph{becomes marginally unfavorable for larger $n = 4,5,6$ clusters} (Fig \ref{fg:na_cn}, Top). 
The observed decrease in $W_n$ with increasing $\lambda$ is in fact found to arise from a increasingly favorable 
medium response: $\mu^{ex}_{\rm outer}(n) -\mu^{ex}_{\alpha}(0)$  decreases as $\lambda$ increases (Fig \ref{fg:na_cn}, Bottom). This
feature emerges because the coordination shell solvent becomes insensitive to the presence of the bulk 
for large coordination radii.  It is this feature that compensates for the increase in $\xi_{\rm aq}(n)$ and leads to the observed
decrease in $W_n$.

\subsection{Cluster variation and optimal coordination states}

\begin{figure}
\includegraphics{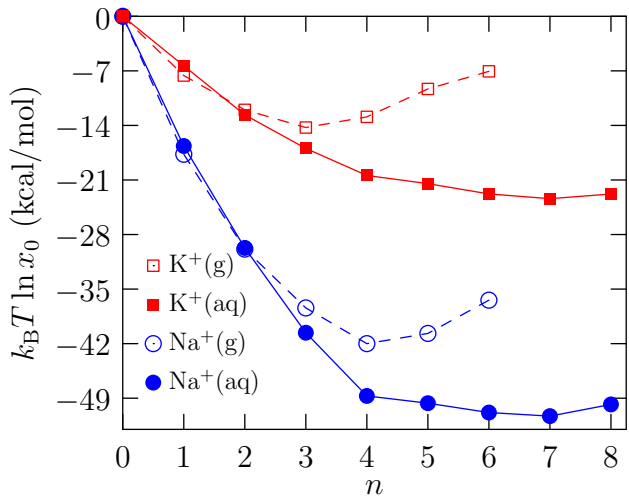}
\caption{A maximum term approximation to the local chemical contribution (Eq.~\ref{eq:lnx0}), $\ln x_0 \approx \beta \beta W_n - \ln p_n/p_0$. For K$^+$, neglecting the bulk medium in the local ion water clustering reaction leads to the identification of the $n=3$ coordination state as the optimal  (or most probable) coordination state.  Account for the medium indicates that $n=7$ is the most probable coordination state  (cf.\ Fig.~7 in Ref.~\onlinecite{merchant:jcp09}).  For Na$^+$, in the presence (absence) of the medium $n=7$ ($n=4$) states are indicated to be optimal (cf.\ Fig.~3 in Ref.~\onlinecite{merchant:jcp09}).}\label{fg:clusV} 
\end{figure}

Figure~\ref{fg:clusV} shows the local chemical contribution (Eq.~\ref{eq:lnx0}) to the hydration free energy; a maximum term 
approximation has been  used. 
Note that Eq.~\ref{eq:lnx0} can be rewritten as \cite{merchant:jcp09}
\begin{eqnarray*}
\ln x_0 = -\ln ( 1 + \sum_{n \geq 1} e^{-\beta W_n} \frac{p_n}{p_0}) \, ,
\end{eqnarray*}
 where $p_n$ is the probability of observing $n$ solvent molecules in the coordination sphere in the absence of the solute. Thus
 the maximum term approximation  is simply $\ln x_0 \approx W_n - \ln p_n/p_0$. When the bulk medium is present, 
 $\xi_{\rm aq}(n)$ defines the local contribution to $W_n$ and when it is absent, $\xi_{\rm g}(n)$ defines the local contribution. 
 In both cases, we use the same long-range contribution $\mu^{ex}_{\rm outer}(n) -\mu^{ex}_{\alpha}(0)$ to $W_n$. 
 
For K$^+$,  Fig.\ \ref{fg:clusV} shows that neglecting the role of the bulk medium in the local ion-water clustering 
shifts the predicted most probable coordination state to $n=3$, whereas the value obtained in simulations is $n=7$ for $\lambda = 3.7$~{\AA}: within the maximum term approximation, $\mu^{\rm ex}_{\rm K^+}$ is a minimum for $n=3$ ($n=7$) in the absence (presence) of the medium. For Na$^+$, neglecting the medium suggests an optimal coordination state of $n=4$, whereas in the presence of the bulk, a value of $n=6$ is predicted. The value obtained in simulations\cite{merchant:jcp09} is $n=6$. For both Na$^+$ and K$^+$, any discrepancy between the optimal coordination predicted using cluster variation and those observed in simulations is well within the uncertainty in the calculations. 
 
 Earlier\cite{merchant:jcp09} we had defined dominant hydration structures on the basis of how an increment in the
coordination number contributes to $k_{\rm B}T \ln x_0$, the local chemical contribution to hydration. If in going from $n\rightarrow n+1$, the contribution to $k_{\rm B}T \ln x_0$ (Eq.~\ref{eq:lnx0}) were only of the order of thermal energies (and substantially smaller than the contribution obtained in going from $n-1 \rightarrow n$), then the 
 $n$-coordinate state was regarded as dominant.  Since this definition is closely tied to the local interactions, it is expected to be
insensitive to the definition of any chemically reasonable coordination radii. Fig.~\ref{fg:clusV} shows that past $n=4$ for both Na$^+$ and K$^+$, the change in $k_{\rm B}T \ln x_0$ is only modest. On this basis we would conclude that for the potential model used here, $n=4$ is the dominant coordination state for Na$^+$ and K$^+$, as was found earlier \cite{merchant:jcp09}.

\section{Concluding discussions}

Effort has focused on using thermochemical data obtained either experimentally or from quantum chemical calculations from gas phase clusters to estimate single ion hydration free energies.  An assumption often implicit in these studies is that the configurations of the clusters in the gas phase are similar to their aqueous counterparts.  Our results show that  this assumption is only true for 
small cluster sizes. For the Na$^+$, K$^+$, and F$^-$ ions considered here, clusters with less than three water molecules satisfy this
requirement.  For larger clusters, the external medium starts to influence the local ion-water interaction. The external
medium stabilizes ion-water cluster configurations that  are \emph{better packed}  around the ion, such that 
the local ion-solvent interaction contribution to the free energy of cluster formation is more favorable in the presence of the medium than without.  For higher coordination states of the ion, and, more generally, for the coordination structure and thermodynamics of 
soft ions, our results show that the medium will play a sizable role in the coordination structure and thermodynamics of the hydrated ion. 
For these cases, accounting for the molecular characteristics of the bulk medium on the ion-water cluster is 
important in inferring the structure and thermodynamics of the hydrated ion. 

\appendix
\section{}\label{sc:appA}

In Ref.\ \onlinecite{merchant:jcp09}, the long-range contribution to $K_n/\tilde{K}_n$ was erroneously left out from the equations. But 
these contributions were all correctly considered in the numerical work and hence no result is affected. Since presenting these corrections also provides a helpful alternative perspective on the equilibrium constants appearing in the quasichemical theory, we note those corrections here. 

Eqs.~10 in Ref.\ \onlinecite{merchant:jcp09} should be rewritten as 
\begin{subequations}
\label{eq:knpf}
\begin{eqnarray}
K_n & =  & \frac{\gamma_n}{n!  z_w^n} \int_v \ldots \int_v e^{-\beta U_{n+1}} e^{-\beta \phi(\mathcal R^n, \beta)} d\mathcal{R}^n \label{eq:error}\\
\tilde{K}_n & = & \frac{1}{n!  z_w^n} \int_v \ldots \int_v e^{-\beta U_{n}} e^{-\beta \phi(\mathcal R^n, \beta)} d\mathcal{R}^n  \; ,
\end{eqnarray}
\end{subequations}
where $\gamma_n = e^{-\beta ( \mu^{ex}_{\rm outer}(n) -\mu^{ex}_{\alpha}(0))}$ (Eq.~\ref{eq:KnbyKnt2}) and $z_w$ is the configurational
partition function of a  water molecules.  $U_{n+1}$ is the potential energy of the ion plus $n$  
solvent molecules within the coordination volume and $U_n$ is the potential energy of the solvent molecules in the absence of the ions. 
$\phi(\mathcal R^n, \beta)$ is the field of the bulk medium on the $n$ solvent ligands in the coordination volume in the absence of the
ion. Eq.~\ref{eq:error} includes the factor $\gamma_n$ that was missing in Eq.~10a in Ref.\ \onlinecite{merchant:jcp09}.

Based on the above definitions, we can show that 
\begin{eqnarray}
(\frac{x_0}{p_0})^{-1} &  = &  \sum_{n} \frac{K_n}{\tilde{K}_n} p_n \nonumber \\  
 & = & \sum_{n} \gamma_n \langle e^{-\beta \Delta U} | n\rangle_0 p_n  \, ,
 \label{eq:winsmall}
 \end{eqnarray}
where once again the factor $\gamma_n$ was left out in Ref.\ \onlinecite{merchant:jcp09}. $\Delta U = U_{n+1} - U_n$ is the interaction  energy of the ion with the $n$ solvent molecules within the coordination sphere. Note that $\xi_{\rm aq}(n) = -\beta\ln \langle e^{-\beta \Delta U} | n\rangle_0$ (Eq.~\ref{eq:KnbyKnt1}).   Eq.~\ref{eq:winsmall} includes the factor $\gamma_n$ that was missing
in Eq.~11, Ref.\  \onlinecite{merchant:jcp09}.

Finally, we note that Eq.~A7 in  Ref.\  \onlinecite{merchant:jcp09} is properly $\mu^{\rm ex}_{\rm outer}(n)$ and not $\mu^{\rm ex}_{\rm outer}(0)$ as was indicated.


\newpage
\begin{thebibliography}{60}
\expandafter\ifx\csname natexlab\endcsname\relax\def\natexlab#1{#1}\fi
\expandafter\ifx\csname bibnamefont\endcsname\relax
  \def\bibnamefont#1{#1}\fi
\expandafter\ifx\csname bibfnamefont\endcsname\relax
  \def\bibfnamefont#1{#1}\fi
\expandafter\ifx\csname citenamefont\endcsname\relax
  \def\citenamefont#1{#1}\fi
\expandafter\ifx\csname url\endcsname\relax
  \def\url#1{\texttt{#1}}\fi
\expandafter\ifx\csname urlprefix\endcsname\relax\def\urlprefix{URL }\fi
\providecommand{\bibinfo}[2]{#2}
\providecommand{\eprint}[2][]{\url{#2}}

\bibitem[{\citenamefont{Castleman~Jr and Bowen~Jr}(1996)}]{castleman:jpc96}
\bibinfo{author}{\bibfnamefont{A.}~\bibnamefont{Castleman~Jr}}
  \bibnamefont{and} \bibinfo{author}{\bibfnamefont{K.~H.}
  \bibnamefont{Bowen~Jr}}, \bibinfo{journal}{J. Phys. Chem.}
  \textbf{\bibinfo{volume}{100}}, \bibinfo{pages}{12911}
  (\bibinfo{year}{1996}).

\bibitem[{\citenamefont{Jungwirth and Tobias}(2006)}]{jung:chemrev06}
\bibinfo{author}{\bibfnamefont{P.}~\bibnamefont{Jungwirth}} \bibnamefont{and}
  \bibinfo{author}{\bibfnamefont{D.}~\bibnamefont{Tobias}},
  \bibinfo{journal}{Chem. Rev.} \textbf{\bibinfo{volume}{106}},
  \bibinfo{pages}{1259} (\bibinfo{year}{2006}).

\bibitem[{\citenamefont{Beyer}(2007)}]{beyer:massspec07}
\bibinfo{author}{\bibfnamefont{M.~K.} \bibnamefont{Beyer}},
  \bibinfo{journal}{Chem. Rev.} \textbf{\bibinfo{volume}{26}},
  \bibinfo{pages}{517} (\bibinfo{year}{2007}).

\bibitem[{\citenamefont{Bondybey and Beyer}(2002)}]{bondy:irevpc02}
\bibinfo{author}{\bibfnamefont{V.~E.} \bibnamefont{Bondybey}} \bibnamefont{and}
  \bibinfo{author}{\bibfnamefont{M.~K.} \bibnamefont{Beyer}},
  \bibinfo{journal}{Intl. Rev. Phys. Chem.} \textbf{\bibinfo{volume}{21}},
  \bibinfo{pages}{277} (\bibinfo{year}{2002}).

\bibitem[{\citenamefont{Robertson and Johnson}(2003)}]{robert:annrev03}
\bibinfo{author}{\bibfnamefont{W.~H.} \bibnamefont{Robertson}}
  \bibnamefont{and} \bibinfo{author}{\bibfnamefont{M.~A.}
  \bibnamefont{Johnson}}, \bibinfo{journal}{Ann. Rev. Phys. Chem.}
  \textbf{\bibinfo{volume}{54}}, \bibinfo{pages}{173} (\bibinfo{year}{2003}).

\bibitem[{\citenamefont{Kebarle}(1977)}]{kebarle:anrevpc77}
\bibinfo{author}{\bibfnamefont{P.}~\bibnamefont{Kebarle}},
  \bibinfo{journal}{Ann. Rev. Phys. Chem.} \textbf{\bibinfo{volume}{28}},
  \bibinfo{pages}{445} (\bibinfo{year}{1977}).

\bibitem[{\citenamefont{Niedner-Schatteburg and
  Bondybey}(2000)}]{niedner:chemrev00}
\bibinfo{author}{\bibfnamefont{G.}~\bibnamefont{Niedner-Schatteburg}}
  \bibnamefont{and} \bibinfo{author}{\bibfnamefont{V.~E.}
  \bibnamefont{Bondybey}}, \bibinfo{journal}{Chem. Rev.}
  \textbf{\bibinfo{volume}{100}}, \bibinfo{pages}{4059} (\bibinfo{year}{2000}).

\bibitem[{\citenamefont{Keutsch and Saykally}(2001)}]{keutsch:pnas01}
\bibinfo{author}{\bibfnamefont{F.~N.} \bibnamefont{Keutsch}} \bibnamefont{and}
  \bibinfo{author}{\bibfnamefont{R.~J.} \bibnamefont{Saykally}},
  \bibinfo{journal}{Proc. Natl. Acad. Sc. USA} \textbf{\bibinfo{volume}{98}},
  \bibinfo{pages}{10533} (\bibinfo{year}{2001}).

\bibitem[{\citenamefont{Coe}(2001)}]{coe:irevpc01}
\bibinfo{author}{\bibfnamefont{J.}~\bibnamefont{Coe}}, \bibinfo{journal}{Intl.
  Rev. Phys. Chem.} \textbf{\bibinfo{volume}{20}}, \bibinfo{pages}{33}
  (\bibinfo{year}{2001}).

\bibitem[{\citenamefont{Klots}(1981)}]{klots:jpc81}
\bibinfo{author}{\bibfnamefont{C.~E.} \bibnamefont{Klots}},
  \bibinfo{journal}{J. Phys. Chem.} \textbf{\bibinfo{volume}{85}},
  \bibinfo{pages}{3585} (\bibinfo{year}{1981}).

\bibitem[{\citenamefont{Tissandier et~al.}(1998)\citenamefont{Tissandier,
  Cowen, Feng, Gundlach, Cohen, Earhart, Coe, and Tuttle~Jr}}]{coe:jpca98}
\bibinfo{author}{\bibfnamefont{M.~D.} \bibnamefont{Tissandier}},
  \bibinfo{author}{\bibfnamefont{K.~A.} \bibnamefont{Cowen}},
  \bibinfo{author}{\bibfnamefont{W.~Y.} \bibnamefont{Feng}},
  \bibinfo{author}{\bibfnamefont{E.}~\bibnamefont{Gundlach}},
  \bibinfo{author}{\bibfnamefont{M.~H.} \bibnamefont{Cohen}},
  \bibinfo{author}{\bibfnamefont{A.~D.} \bibnamefont{Earhart}},
  \bibinfo{author}{\bibfnamefont{J.~V.} \bibnamefont{Coe}}, \bibnamefont{and}
  \bibinfo{author}{\bibfnamefont{T.~R.} \bibnamefont{Tuttle~Jr}},
  \bibinfo{journal}{J. Phys. Chem. A} \textbf{\bibinfo{volume}{102}},
  \bibinfo{pages}{7787} (\bibinfo{year}{1998}).

\bibitem[{\citenamefont{Xantheas}(1995)}]{xantheas:jacs95}
\bibinfo{author}{\bibfnamefont{S.~S.} \bibnamefont{Xantheas}},
  \bibinfo{journal}{J. Am. Chem. Soc.} \textbf{\bibinfo{volume}{117}},
  \bibinfo{pages}{10373} (\bibinfo{year}{1995}).

\bibitem[{\citenamefont{Novoa et~al.}(1997)\citenamefont{Novoa, Mota, Valle,
  and Planas}}]{novoa:jpca97}
\bibinfo{author}{\bibfnamefont{J.~J.} \bibnamefont{Novoa}},
  \bibinfo{author}{\bibfnamefont{F.}~\bibnamefont{Mota}},
  \bibinfo{author}{\bibfnamefont{C.}~\bibnamefont{Valle}}, \bibnamefont{and}
  \bibinfo{author}{\bibfnamefont{M.}~\bibnamefont{Planas}},
  \bibinfo{journal}{J. Phys. Chem. A} \textbf{\bibinfo{volume}{101}},
  \bibinfo{pages}{7842} (\bibinfo{year}{1997}).

\bibitem[{\citenamefont{Topol et~al.}(1999)\citenamefont{Topol, Tawa, Burt, and
  Rashin}}]{topol:jcp99}
\bibinfo{author}{\bibfnamefont{I.~A.} \bibnamefont{Topol}},
  \bibinfo{author}{\bibfnamefont{G.}~\bibnamefont{Tawa}},
  \bibinfo{author}{\bibfnamefont{S.}~\bibnamefont{Burt}}, \bibnamefont{and}
  \bibinfo{author}{\bibfnamefont{A.~A.} \bibnamefont{Rashin}},
  \bibinfo{journal}{J. Chem. Phys.} \textbf{\bibinfo{volume}{111}},
  \bibinfo{pages}{10998} (\bibinfo{year}{1999}).

\bibitem[{\citenamefont{Zhan and Dixon}(2001)}]{zhan:jcpa01}
\bibinfo{author}{\bibfnamefont{C.~G.} \bibnamefont{Zhan}} \bibnamefont{and}
  \bibinfo{author}{\bibfnamefont{D.~A.} \bibnamefont{Dixon}},
  \bibinfo{journal}{J. Phys. Chem. A} \textbf{\bibinfo{volume}{105}},
  \bibinfo{pages}{11534} (\bibinfo{year}{2001}).

\bibitem[{\citenamefont{Pliego~Jr. and Riveros}(2001)}]{pliego:jcpa01}
\bibinfo{author}{\bibfnamefont{J.~R.} \bibnamefont{Pliego~Jr.}}
  \bibnamefont{and} \bibinfo{author}{\bibfnamefont{J.~M.}
  \bibnamefont{Riveros}}, \bibinfo{journal}{J. Phys. Chem. A}
  \textbf{\bibinfo{volume}{105}}, \bibinfo{pages}{7241} (\bibinfo{year}{2001}).

\bibitem[{\citenamefont{Bryantsev et~al.}(2008)\citenamefont{Bryantsev, Diallo,
  and Goddard~III}}]{brya:jpcb08}
\bibinfo{author}{\bibfnamefont{V.~S.} \bibnamefont{Bryantsev}},
  \bibinfo{author}{\bibfnamefont{M.~S.} \bibnamefont{Diallo}},
  \bibnamefont{and} \bibinfo{author}{\bibfnamefont{W.~A.}
  \bibnamefont{Goddard~III}}, \bibinfo{journal}{J. Phys. Chem. B}
  \textbf{\bibinfo{volume}{112}}, \bibinfo{pages}{9709} (\bibinfo{year}{2008}).

\bibitem[{\citenamefont{Kelly et~al.}(2006)\citenamefont{Kelly, Cramer, and
  Truhlar}}]{kelly:jpcb06}
\bibinfo{author}{\bibfnamefont{C.~P.} \bibnamefont{Kelly}},
  \bibinfo{author}{\bibfnamefont{C.~J.} \bibnamefont{Cramer}},
  \bibnamefont{and} \bibinfo{author}{\bibfnamefont{D.~G.}
  \bibnamefont{Truhlar}}, \bibinfo{journal}{J. Phys. Chem. B}
  \textbf{\bibinfo{volume}{110}}, \bibinfo{pages}{16066}
  (\bibinfo{year}{2006}).

\bibitem[{\citenamefont{Bustamante et~al.}(2011)\citenamefont{Bustamante,
  Valencia, and Castro}}]{busta:jpca11}
\bibinfo{author}{\bibfnamefont{M.}~\bibnamefont{Bustamante}},
  \bibinfo{author}{\bibfnamefont{I.}~\bibnamefont{Valencia}}, \bibnamefont{and}
  \bibinfo{author}{\bibfnamefont{M.}~\bibnamefont{Castro}},
  \bibinfo{journal}{J. Phys. Chem. A} \textbf{\bibinfo{volume}{115}},
  \bibinfo{pages}{4115} (\bibinfo{year}{2011}).

\bibitem[{\citenamefont{Pratt and La{V}iolette}(1998)}]{lrp:mp98}
\bibinfo{author}{\bibfnamefont{L.~R.} \bibnamefont{Pratt}} \bibnamefont{and}
  \bibinfo{author}{\bibfnamefont{R.~A.} \bibnamefont{La{V}iolette}},
  \bibinfo{journal}{Mol. Phys.} \textbf{\bibinfo{volume}{94}},
  \bibinfo{pages}{909 } (\bibinfo{year}{1998}).

\bibitem[{\citenamefont{Pratt and Rempe}(1999)}]{lrp:ES99}
\bibinfo{author}{\bibfnamefont{L.~R.} \bibnamefont{Pratt}} \bibnamefont{and}
  \bibinfo{author}{\bibfnamefont{S.~B.} \bibnamefont{Rempe}}, in
  \emph{\bibinfo{booktitle}{Simulation and Theory of Electrostatic Interactions
  in Solution. Computational Chemistry, Biophysics, and Aqueous Solutions}},
  edited by \bibinfo{editor}{\bibfnamefont{L.~R.} \bibnamefont{Pratt}}
  \bibnamefont{and} \bibinfo{editor}{\bibfnamefont{G.}~\bibnamefont{Hummer}}
  (\bibinfo{publisher}{American Institute of Physics},
  \bibinfo{address}{Melville, NY}, \bibinfo{year}{1999}), vol.
  \bibinfo{volume}{492} of \emph{\bibinfo{series}{AIP Conference Proceedings}},
  pp. \bibinfo{pages}{172--201}.

\bibitem[{\citenamefont{Paulaitis and Pratt}(2002)}]{pauli:advpc02}
\bibinfo{author}{\bibfnamefont{M.~E.} \bibnamefont{Paulaitis}}
  \bibnamefont{and} \bibinfo{author}{\bibfnamefont{L.}~\bibnamefont{Pratt}},
  \bibinfo{journal}{Adv. Prot. Chem.} \textbf{\bibinfo{volume}{62}},
  \bibinfo{pages}{283} (\bibinfo{year}{2002}).

\bibitem[{\citenamefont{Beck et~al.}(2006)\citenamefont{Beck, Paulaitis, and
  Pratt}}]{lrp:book}
\bibinfo{author}{\bibfnamefont{T.~L.} \bibnamefont{Beck}},
  \bibinfo{author}{\bibfnamefont{M.~E.} \bibnamefont{Paulaitis}},
  \bibnamefont{and} \bibinfo{author}{\bibfnamefont{L.~R.} \bibnamefont{Pratt}},
  \emph{\bibinfo{title}{The potential distribution theorem and models of
  molecular solutions}} (\bibinfo{publisher}{Cambridge University Press},
  \bibinfo{year}{2006}).

\bibitem[{\citenamefont{Pratt and Asthagiri}(2007)}]{lrp:cpms}
\bibinfo{author}{\bibfnamefont{L.~R.} \bibnamefont{Pratt}} \bibnamefont{and}
  \bibinfo{author}{\bibfnamefont{D.}~\bibnamefont{Asthagiri}}, in
  \emph{\bibinfo{booktitle}{Free energy calculations: {Theory} and applications
  in chemistry and biology}}, edited by
  \bibinfo{editor}{\bibfnamefont{C.}~\bibnamefont{Chipot}} \bibnamefont{and}
  \bibinfo{editor}{\bibfnamefont{A.}~\bibnamefont{Pohorille}}
  (\bibinfo{publisher}{Springer}, \bibinfo{year}{2007}),
  vol.~\bibinfo{volume}{86} of \emph{\bibinfo{series}{Springer series in
  {Chemical Physics}}}, chap.~\bibinfo{chapter}{9}, pp.
  \bibinfo{pages}{323--351}.

\bibitem[{\citenamefont{Rempe et~al.}(2000)\citenamefont{Rempe, Pratt, and
  Hummer}}]{rempe:jacs00}
\bibinfo{author}{\bibfnamefont{S.~B.} \bibnamefont{Rempe}},
  \bibinfo{author}{\bibfnamefont{L.}~\bibnamefont{Pratt}}, \bibnamefont{and}
  \bibinfo{author}{\bibfnamefont{G.}~\bibnamefont{Hummer}},
  \bibinfo{journal}{J. Am. Chem. Soc.} \textbf{\bibinfo{volume}{122}},
  \bibinfo{pages}{966} (\bibinfo{year}{2000}).

\bibitem[{\citenamefont{Rempe and Pratt}(2001)}]{rempe:fpe01}
\bibinfo{author}{\bibfnamefont{S.~B.} \bibnamefont{Rempe}} \bibnamefont{and}
  \bibinfo{author}{\bibfnamefont{L.~R.} \bibnamefont{Pratt}},
  \bibinfo{journal}{Fluid Phase Equilibria} \textbf{\bibinfo{volume}{183-184}},
  \bibinfo{pages}{121} (\bibinfo{year}{2001}).

\bibitem[{\citenamefont{Grabowski et~al.}(2002)\citenamefont{Grabowski,
  Riccardi, Gomez, Asthagiri, and Pratt}}]{grab:jpca02}
\bibinfo{author}{\bibfnamefont{P.}~\bibnamefont{Grabowski}},
  \bibinfo{author}{\bibfnamefont{D.}~\bibnamefont{Riccardi}},
  \bibinfo{author}{\bibfnamefont{M.~A.} \bibnamefont{Gomez}},
  \bibinfo{author}{\bibfnamefont{D.}~\bibnamefont{Asthagiri}},
  \bibnamefont{and} \bibinfo{author}{\bibfnamefont{L.}~\bibnamefont{Pratt}},
  \bibinfo{journal}{J. Phys. Chem. A} \textbf{\bibinfo{volume}{106}},
  \bibinfo{pages}{9145} (\bibinfo{year}{2002}).

\bibitem[{\citenamefont{Asthagiri and Pratt}(2003)}]{asthagiri:cpl03}
\bibinfo{author}{\bibfnamefont{D.}~\bibnamefont{Asthagiri}} \bibnamefont{and}
  \bibinfo{author}{\bibfnamefont{L.}~\bibnamefont{Pratt}},
  \bibinfo{journal}{Chem. Phys. Lett.} \textbf{\bibinfo{volume}{371}},
  \bibinfo{pages}{613} (\bibinfo{year}{2003}).

\bibitem[{\citenamefont{Asthagiri
  et~al.}(2003{\natexlab{a}})\citenamefont{Asthagiri, Pratt, Kress, and
  Gomez}}]{asthagiri:cpl032}
\bibinfo{author}{\bibfnamefont{D.}~\bibnamefont{Asthagiri}},
  \bibinfo{author}{\bibfnamefont{L.}~\bibnamefont{Pratt}},
  \bibinfo{author}{\bibfnamefont{J.~D.} \bibnamefont{Kress}}, \bibnamefont{and}
  \bibinfo{author}{\bibfnamefont{M.~A.} \bibnamefont{Gomez}},
  \bibinfo{journal}{Chem. Phys. Lett.} \textbf{\bibinfo{volume}{380}},
  \bibinfo{pages}{530} (\bibinfo{year}{2003}{\natexlab{a}}).

\bibitem[{\citenamefont{Rempe et~al.}(2004)\citenamefont{Rempe, Asthagiri, and
  Pratt}}]{rempe:pccp04}
\bibinfo{author}{\bibfnamefont{S.~B.} \bibnamefont{Rempe}},
  \bibinfo{author}{\bibfnamefont{D.}~\bibnamefont{Asthagiri}},
  \bibnamefont{and} \bibinfo{author}{\bibfnamefont{L.}~\bibnamefont{Pratt}},
  \bibinfo{journal}{Phys. Chem. Chem. Phys.} \textbf{\bibinfo{volume}{6}},
  \bibinfo{pages}{1966} (\bibinfo{year}{2004}).

\bibitem[{\citenamefont{Asthagiri
  et~al.}(2003{\natexlab{b}})\citenamefont{Asthagiri, Pratt, and
  Ashbaugh}}]{asthagiri:jcp03}
\bibinfo{author}{\bibfnamefont{D.}~\bibnamefont{Asthagiri}},
  \bibinfo{author}{\bibfnamefont{L.}~\bibnamefont{Pratt}}, \bibnamefont{and}
  \bibinfo{author}{\bibfnamefont{H.~S.} \bibnamefont{Ashbaugh}},
  \bibinfo{journal}{J. Chem. Phys.} \textbf{\bibinfo{volume}{119}},
  \bibinfo{pages}{2702} (\bibinfo{year}{2003}{\natexlab{b}}).

\bibitem[{\citenamefont{Asthagiri
  et~al.}(2004{\natexlab{a}})\citenamefont{Asthagiri, Pratt, Paulaitis, and
  Rempe}}]{lrp:zn04}
\bibinfo{author}{\bibfnamefont{D.}~\bibnamefont{Asthagiri}},
  \bibinfo{author}{\bibfnamefont{L.~R.} \bibnamefont{Pratt}},
  \bibinfo{author}{\bibfnamefont{M.~E.} \bibnamefont{Paulaitis}},
  \bibnamefont{and} \bibinfo{author}{\bibfnamefont{S.~B.} \bibnamefont{Rempe}},
  \bibinfo{journal}{J. Am. Chem. Soc.} \textbf{\bibinfo{volume}{126}},
  \bibinfo{pages}{1285} (\bibinfo{year}{2004}{\natexlab{a}}).

\bibitem[{\citenamefont{Asthagiri
  et~al.}(2003{\natexlab{c}})\citenamefont{Asthagiri, Pratt, and
  Kress}}]{asthagiri:waterpre}
\bibinfo{author}{\bibfnamefont{D.}~\bibnamefont{Asthagiri}},
  \bibinfo{author}{\bibfnamefont{L.~R.} \bibnamefont{Pratt}}, \bibnamefont{and}
  \bibinfo{author}{\bibfnamefont{J.~D.} \bibnamefont{Kress}},
  \bibinfo{journal}{Phys. Rev. E} \textbf{\bibinfo{volume}{68}},
  \bibinfo{pages}{041505} (\bibinfo{year}{2003}{\natexlab{c}}).

\bibitem[{\citenamefont{Grossman et~al.}(2004)\citenamefont{Grossman,
  Schwegler, Draeger, Gygi, and Galli}}]{schwegler:jcp04a}
\bibinfo{author}{\bibfnamefont{J.~C.} \bibnamefont{Grossman}},
  \bibinfo{author}{\bibfnamefont{E.}~\bibnamefont{Schwegler}},
  \bibinfo{author}{\bibfnamefont{E.~W.} \bibnamefont{Draeger}},
  \bibinfo{author}{\bibfnamefont{F.}~\bibnamefont{Gygi}}, \bibnamefont{and}
  \bibinfo{author}{\bibfnamefont{G.}~\bibnamefont{Galli}}, \bibinfo{journal}{J.
  Chem. Phys.} \textbf{\bibinfo{volume}{120}}, \bibinfo{pages}{300}
  (\bibinfo{year}{2004}).

\bibitem[{\citenamefont{Asthagiri et~al.}(2005)\citenamefont{Asthagiri, Pratt,
  and Kress}}]{asthagiri:hpnas}
\bibinfo{author}{\bibfnamefont{D.}~\bibnamefont{Asthagiri}},
  \bibinfo{author}{\bibfnamefont{L.~R.} \bibnamefont{Pratt}}, \bibnamefont{and}
  \bibinfo{author}{\bibfnamefont{J.~D.} \bibnamefont{Kress}},
  \bibinfo{journal}{Proc. Natl. Acad. Sc. USA} \textbf{\bibinfo{volume}{102}},
  \bibinfo{pages}{6704} (\bibinfo{year}{2005}).

\bibitem[{\citenamefont{Asthagiri
  et~al.}(2004{\natexlab{b}})\citenamefont{Asthagiri, Pratt, Kress, and
  Gomez}}]{asthagiri:hopnas}
\bibinfo{author}{\bibfnamefont{D.}~\bibnamefont{Asthagiri}},
  \bibinfo{author}{\bibfnamefont{L.~R.} \bibnamefont{Pratt}},
  \bibinfo{author}{\bibfnamefont{J.~D.} \bibnamefont{Kress}}, \bibnamefont{and}
  \bibinfo{author}{\bibfnamefont{M.~A.} \bibnamefont{Gomez}},
  \bibinfo{journal}{Proc. Natl. Acad. Sc. USA} \textbf{\bibinfo{volume}{101}},
  \bibinfo{pages}{7229} (\bibinfo{year}{2004}{\natexlab{b}}).

\bibitem[{\citenamefont{Voth}(2006)}]{voth:acc06}
\bibinfo{author}{\bibfnamefont{G.~A.} \bibnamefont{Voth}},
  \bibinfo{journal}{Acc. Chem. Res.} \textbf{\bibinfo{volume}{39}},
  \bibinfo{pages}{143} (\bibinfo{year}{2006}).

\bibitem[{\citenamefont{Tuckerman et~al.}(2006)\citenamefont{Tuckerman,
  Chandra, and Marx}}]{tuckerman:acc06}
\bibinfo{author}{\bibfnamefont{M.~E.} \bibnamefont{Tuckerman}},
  \bibinfo{author}{\bibfnamefont{A.}~\bibnamefont{Chandra}}, \bibnamefont{and}
  \bibinfo{author}{\bibfnamefont{D.}~\bibnamefont{Marx}},
  \bibinfo{journal}{Acc. Chem. Res.} \textbf{\bibinfo{volume}{39}},
  \bibinfo{pages}{151} (\bibinfo{year}{2006}).

\bibitem[{\citenamefont{Varma and Rempe}(2006)}]{rempe:bc06}
\bibinfo{author}{\bibfnamefont{S.}~\bibnamefont{Varma}} \bibnamefont{and}
  \bibinfo{author}{\bibfnamefont{S.~B.} \bibnamefont{Rempe}},
  \bibinfo{journal}{Biophys. Chem.} \textbf{\bibinfo{volume}{124}},
  \bibinfo{pages}{192} (\bibinfo{year}{2006}).

\bibitem[{\citenamefont{Whitfield et~al.}(2007)\citenamefont{Whitfield, Varma,
  Harder, Lamoureux, Rempe, and Roux}}]{whitfield:jctc07}
\bibinfo{author}{\bibfnamefont{T.~W.} \bibnamefont{Whitfield}},
  \bibinfo{author}{\bibfnamefont{S.}~\bibnamefont{Varma}},
  \bibinfo{author}{\bibfnamefont{E.}~\bibnamefont{Harder}},
  \bibinfo{author}{\bibfnamefont{G.}~\bibnamefont{Lamoureux}},
  \bibinfo{author}{\bibfnamefont{S.~B.} \bibnamefont{Rempe}}, \bibnamefont{and}
  \bibinfo{author}{\bibfnamefont{B.}~\bibnamefont{Roux}}, \bibinfo{journal}{J.
  Chem. Theory Comput.} \textbf{\bibinfo{volume}{3}}, \bibinfo{pages}{2068}
  (\bibinfo{year}{2007}).

\bibitem[{\citenamefont{Xantheas}(1996)}]{xantheas:jpc96}
\bibinfo{author}{\bibfnamefont{S.~S.} \bibnamefont{Xantheas}},
  \bibinfo{journal}{J. Phys. Chem.} \textbf{\bibinfo{volume}{100}},
  \bibinfo{pages}{9703} (\bibinfo{year}{1996}).

\bibitem[{\citenamefont{Merchant and Asthagiri}(2009)}]{merchant:jcp09}
\bibinfo{author}{\bibfnamefont{S.}~\bibnamefont{Merchant}} \bibnamefont{and}
  \bibinfo{author}{\bibfnamefont{D.}~\bibnamefont{Asthagiri}},
  \bibinfo{journal}{J. Chem. Phys.} \textbf{\bibinfo{volume}{130}},
  \bibinfo{pages}{195102} (\bibinfo{year}{2009}).

\bibitem[{\citenamefont{Paliwal et~al.}(2006)\citenamefont{Paliwal, Asthagiri,
  Pratt, Ashbaugh, and Paulaitis}}]{paliwal:jcp06}
\bibinfo{author}{\bibfnamefont{A.}~\bibnamefont{Paliwal}},
  \bibinfo{author}{\bibfnamefont{D.}~\bibnamefont{Asthagiri}},
  \bibinfo{author}{\bibfnamefont{L.~R.} \bibnamefont{Pratt}},
  \bibinfo{author}{\bibfnamefont{H.~S.} \bibnamefont{Ashbaugh}},
  \bibnamefont{and} \bibinfo{author}{\bibfnamefont{M.~E.}
  \bibnamefont{Paulaitis}}, \bibinfo{journal}{J. Chem. Phys.}
  \textbf{\bibinfo{volume}{124}}, \bibinfo{pages}{224502}
  (\bibinfo{year}{2006}).

\bibitem[{\citenamefont{Pratt and Laviolette}(1998)}]{pratt:mp98}
\bibinfo{author}{\bibfnamefont{L.}~\bibnamefont{Pratt}} \bibnamefont{and}
  \bibinfo{author}{\bibfnamefont{R.~A.} \bibnamefont{Laviolette}},
  \bibinfo{journal}{Mol Phys} \textbf{\bibinfo{volume}{94}},
  \bibinfo{pages}{909} (\bibinfo{year}{1998}).

\bibitem[{\citenamefont{Pratt et~al.}(2001)\citenamefont{Pratt, LaViolette,
  Gomez, and Gentile}}]{lrp:jpcb01}
\bibinfo{author}{\bibfnamefont{L.~R.} \bibnamefont{Pratt}},
  \bibinfo{author}{\bibfnamefont{R.~A.} \bibnamefont{LaViolette}},
  \bibinfo{author}{\bibfnamefont{M.~A.} \bibnamefont{Gomez}}, \bibnamefont{and}
  \bibinfo{author}{\bibfnamefont{M.~E.} \bibnamefont{Gentile}},
  \bibinfo{journal}{J. Phys. Chem. B} \textbf{\bibinfo{volume}{105}},
  \bibinfo{pages}{11662 } (\bibinfo{year}{2001}).

\bibitem[{\citenamefont{Pratt and Ashbaugh}(2003)}]{lrp:hspre}
\bibinfo{author}{\bibfnamefont{L.~R.} \bibnamefont{Pratt}} \bibnamefont{and}
  \bibinfo{author}{\bibfnamefont{H.~S.} \bibnamefont{Ashbaugh}},
  \bibinfo{journal}{Phys. Rev. E} \textbf{\bibinfo{volume}{68}},
  \bibinfo{pages}{021505} (\bibinfo{year}{2003}).

\bibitem[{\citenamefont{Metropolis et~al.}(1953)\citenamefont{Metropolis,
  Rosenbluth, Rosenbluth, Teller, and Teller}}]{metropolis:jcp53}
\bibinfo{author}{\bibfnamefont{N.}~\bibnamefont{Metropolis}},
  \bibinfo{author}{\bibfnamefont{A.~W.} \bibnamefont{Rosenbluth}},
  \bibinfo{author}{\bibfnamefont{M.~N.} \bibnamefont{Rosenbluth}},
  \bibinfo{author}{\bibfnamefont{A.~H.} \bibnamefont{Teller}},
  \bibnamefont{and} \bibinfo{author}{\bibfnamefont{E.}~\bibnamefont{Teller}},
  \bibinfo{journal}{J. Chem. Phys.} \textbf{\bibinfo{volume}{21}},
  \bibinfo{pages}{1087} (\bibinfo{year}{1953}).

\bibitem[{\citenamefont{Allen and Tildesley}(1987)}]{allen}
\bibinfo{author}{\bibfnamefont{M.~P.} \bibnamefont{Allen}} \bibnamefont{and}
  \bibinfo{author}{\bibfnamefont{D.~J.} \bibnamefont{Tildesley}},
  \emph{\bibinfo{title}{Computer simulation of liquids}}
  (\bibinfo{publisher}{Clarendon Press}, \bibinfo{address}{Oxford},
  \bibinfo{year}{1987}).

\bibitem[{\citenamefont{Berendsen et~al.}(1987)\citenamefont{Berendsen,
  Grigera, and Straatsma}}]{berendsen:jpc87}
\bibinfo{author}{\bibfnamefont{H.~J.~C.} \bibnamefont{Berendsen}},
  \bibinfo{author}{\bibfnamefont{J.~R.} \bibnamefont{Grigera}},
  \bibnamefont{and} \bibinfo{author}{\bibfnamefont{T.~P.}
  \bibnamefont{Straatsma}}, \bibinfo{journal}{J. Phys. Chem.}
  \textbf{\bibinfo{volume}{91}}, \bibinfo{pages}{6269} (\bibinfo{year}{1987}).

\bibitem[{\citenamefont{Hummer et~al.}(1996)\citenamefont{Hummer, Pratt, and
  Garcia}}]{hummer:jpc96}
\bibinfo{author}{\bibfnamefont{G.}~\bibnamefont{Hummer}},
  \bibinfo{author}{\bibfnamefont{L.~R.} \bibnamefont{Pratt}}, \bibnamefont{and}
  \bibinfo{author}{\bibfnamefont{A.~E.} \bibnamefont{Garcia}},
  \bibinfo{journal}{J. Phys. Chem.} \textbf{\bibinfo{volume}{100}},
  \bibinfo{pages}{1206} (\bibinfo{year}{1996}).

\bibitem[{\citenamefont{Hummer et~al.}(1992)\citenamefont{Hummer, Soumpasis,
  and Neumann}}]{hummer:molphys92}
\bibinfo{author}{\bibfnamefont{G.}~\bibnamefont{Hummer}},
  \bibinfo{author}{\bibfnamefont{D.~M.} \bibnamefont{Soumpasis}},
  \bibnamefont{and} \bibinfo{author}{\bibfnamefont{M.}~\bibnamefont{Neumann}},
  \bibinfo{journal}{Mol Phys} \textbf{\bibinfo{volume}{77}},
  \bibinfo{pages}{769} (\bibinfo{year}{1992}).

\bibitem[{\citenamefont{Hummer and Soumpasis}(1994)}]{hummer:physreve94}
\bibinfo{author}{\bibfnamefont{G.}~\bibnamefont{Hummer}} \bibnamefont{and}
  \bibinfo{author}{\bibfnamefont{D.~M.} \bibnamefont{Soumpasis}},
  \bibinfo{journal}{Phys. Rev. E} \textbf{\bibinfo{volume}{49}},
  \bibinfo{pages}{591} (\bibinfo{year}{1994}).

\bibitem[{\citenamefont{Hummer et~al.}(1994)\citenamefont{Hummer, Soumpasis,
  and Neumann}}]{hummer:jpcond94}
\bibinfo{author}{\bibfnamefont{G.}~\bibnamefont{Hummer}},
  \bibinfo{author}{\bibfnamefont{D.~M.} \bibnamefont{Soumpasis}},
  \bibnamefont{and} \bibinfo{author}{\bibfnamefont{M.}~\bibnamefont{Neumann}},
  \bibinfo{journal}{Journal of Physics: Condensed Matter}
  \textbf{\bibinfo{volume}{6}}, \bibinfo{pages}{A141} (\bibinfo{year}{1994}).

\bibitem[{\citenamefont{Hummer et~al.}(1998)\citenamefont{Hummer, Pratt, and
  Garcia}}]{lrp:ionsjpca98}
\bibinfo{author}{\bibfnamefont{G.}~\bibnamefont{Hummer}},
  \bibinfo{author}{\bibfnamefont{L.~R.} \bibnamefont{Pratt}}, \bibnamefont{and}
  \bibinfo{author}{\bibfnamefont{A.~E.} \bibnamefont{Garcia}},
  \bibinfo{journal}{J. Chem. Phys. A} \textbf{\bibinfo{volume}{102}},
  \bibinfo{pages}{7885 } (\bibinfo{year}{1998}).

\bibitem[{\citenamefont{Hummer and Szabo}(1996)}]{hummer:jcp96}
\bibinfo{author}{\bibfnamefont{G.}~\bibnamefont{Hummer}} \bibnamefont{and}
  \bibinfo{author}{\bibfnamefont{A.}~\bibnamefont{Szabo}}, \bibinfo{journal}{J.
  Chem. Phys.} \textbf{\bibinfo{volume}{105}}, \bibinfo{pages}{2004}
  (\bibinfo{year}{1996}).

\bibitem[{\citenamefont{Asthagiri et~al.}(2007)\citenamefont{Asthagiri,
  Ashbaugh, Piryatinski, Paulaitis, and Pratt}}]{Asthagiri:2007p323}
\bibinfo{author}{\bibfnamefont{D.}~\bibnamefont{Asthagiri}},
  \bibinfo{author}{\bibfnamefont{H.~S.} \bibnamefont{Ashbaugh}},
  \bibinfo{author}{\bibfnamefont{A.}~\bibnamefont{Piryatinski}},
  \bibinfo{author}{\bibfnamefont{M.~E.} \bibnamefont{Paulaitis}},
  \bibnamefont{and} \bibinfo{author}{\bibfnamefont{L.~R.} \bibnamefont{Pratt}},
  \bibinfo{journal}{J. Am. Chem. Soc.} \textbf{\bibinfo{volume}{129}},
  \bibinfo{pages}{10133} (\bibinfo{year}{2007}).

\bibitem[{\citenamefont{Bennett}(1976)}]{bennett:jcp76}
\bibinfo{author}{\bibfnamefont{C.~H.} \bibnamefont{Bennett}},
  \bibinfo{journal}{J. Comp. Phys.} \textbf{\bibinfo{volume}{22}},
  \bibinfo{pages}{245} (\bibinfo{year}{1976}).

\bibitem[{\citenamefont{Meot-Ner and Speller}(1986)}]{maut:jpc86}
\bibinfo{author}{\bibfnamefont{M.}~\bibnamefont{Meot-Ner}} \bibnamefont{and}
  \bibinfo{author}{\bibfnamefont{C.~V.} \bibnamefont{Speller}},
  \bibinfo{journal}{J. Phys. Chem.} \textbf{\bibinfo{volume}{90}},
  \bibinfo{pages}{6616} (\bibinfo{year}{1986}).

\bibitem[{\citenamefont{Searles and Kerbale}(1968)}]{searles:jpc68}
\bibinfo{author}{\bibfnamefont{S.~K.} \bibnamefont{Searles}} \bibnamefont{and}
  \bibinfo{author}{\bibfnamefont{P.}~\bibnamefont{Kerbale}},
  \bibinfo{journal}{J. Phys. Chem.} \textbf{\bibinfo{volume}{72}},
  \bibinfo{pages}{742} (\bibinfo{year}{1968}).

\bibitem[{\citenamefont{Castleman et~al.}(1978)\citenamefont{Castleman,
  Holland, Lindsay, and Peterson}}]{castleman:jacs78}
\bibinfo{author}{\bibfnamefont{A.~W.} \bibnamefont{Castleman}},
  \bibinfo{author}{\bibfnamefont{P.~M.} \bibnamefont{Holland}},
  \bibinfo{author}{\bibfnamefont{D.~M.} \bibnamefont{Lindsay}},
  \bibnamefont{and} \bibinfo{author}{\bibfnamefont{K.~I.}
  \bibnamefont{Peterson}}, \bibinfo{journal}{J. Am. Chem. Soc.}
  \textbf{\bibinfo{volume}{100}}, \bibinfo{pages}{6039} (\bibinfo{year}{1978}).

\end{thebibliography}
 \end{document}